\documentclass{svproc}
\usepackage{url}
\usepackage{hyperref}
\usepackage{graphicx}
\usepackage{subfigure}
\usepackage{multirow}
\usepackage{amsmath}

\begin{document}
\mainmatter              
\title{Deconstruct and Reconstruct Dizi Music of the Northern School and the Southern School}
\titlerunning{Deconstruct and Reconstruct Dizi Music}  
%
\author{Yifan Xie \and Rongfeng Li}
\authorrunning{Yifan Xie et al.} 
%
%
\institute{Beijing Key Laboratory of Network System and Network Culture, Beijing University of Posts and Telecommunications, Beijing 100876, China\\
\email{\{yifan.xie, lirongfeng\}@bupt.edu.cn}}

\maketitle              

\begin{abstract}
Today's research on Chinese music technology is mainly focused on three aspects: data collection, music deconstruction, and music reconstruction. In this paper, a general method is proposed to collect Chinese music in the form of numbered musical notation, and a Dizi dataset is collected using this method. Based on the collected Dizi dataset, we conduct research on the Dizi music styles of the Northern school and the Southern School. Characteristics include melody and playing techniques of the two different music styles are deconstructed. A reconstruction example, music style transfer which includes melody transfer and playing techniques transfer is given and audience evaluation is done to evaluate the reconstruction results.  

\end{abstract}
\section{Introduction}
With the continuous development of music technology, more and more researchers are devoted to the exploration of Chinese music technology. These studies mainly focus on three aspects: the collection of Chinese music datasets, the deconstruction of Chinese music, and the reconstruction of Chinese music. Collected datasets include two forms: audio and symbolic scores. Deconstruction refers to data mining from the collected datasets, to find some information about features from music. Reconstruction refers to the creation of new music, new musical forms, and so on.

Of course, there are still many areas worthy of improvement. The current research on Chinese music mainly has the following problems:

\begin{itemize}
	\item In terms of data collection, there has been a relatively standardized and systematic collection method for audio, and there is also a certain scale of Chinese musical instrument database \cite{liang2019constructing}. But in terms of the establishment of a symbolic score database, although some methods have been proposed before, there is still no standard, accurate and fast collection method for the numbered musical scores. For example, Optical Character Recognition technology is a fast method but could lead to many errors due to immaturity. The collection method proposed in \cite{guqin} for Guqin is fast and accurate to some extent, but could not intuitively be represented in the form of numbered musical notation. 
	
	\item In the reconstruction and deconstruction of music, the gap between music and computer science has caused two hands of problems. On the one hand, for researchers in computer science, much research stays at the stage of pure data analysis without an in-depth discussion of the meaning of the music. Simply migrating methods from other fields to Chinese music technology would not bring substantial progress to the research of Chinese music technology. On the other hand, for researchers in music, much research fails to make good use of the powerful tool of computer science. 
	
\end{itemize}

In response to the above problems. Taking Dizi music as an example, we do the following work in this paper:

\begin{itemize}
	\item we propose a general collection method to collect numbered musical scores by typing them using a self-made font. In this way, collected scores can be represented intuitively in the form of numbered musical notation. Then, these scores can be converted into staff easily using a written program. Using this method, we collect the first symbolic dataset of Dizi music. We also make public the dataset\footnote{\href{https://github.com/hrsoup/Dizi\_Dataset}{https://github.com/hrsoup/Dizi\_Dataset}}.
	
	\item Based on the Dizi dataset, we do data mining (deconstruction) include melody deconstruction and playing techniques deconstruction on the Dizi music styles. Playing techniques in Chinese music is much more important than in Western music. Through deconstruction, we not only lay the foundation for the later music reconstruction but also find some interesting phenomena. 
	
	\item Based on the deconstruction results, we give an interesting reconstruction example, music style transfer, which includes both melody transfer and playing techniques transfer. Some audience tests are done to evaluate the transfer results.  
	
\end{itemize}

The code used in this paper can be found online\footnote{\href{https://github.com/hrsoup/CSMT2020\_Code}{https://github.com/hrsoup/CSMT2020\_Code}}. Besides, here is a brief introduction to styles of Dizi music, especially for the styles of the Northern school and the Southern school, which are seen as the research objects in this paper. In the 1950s, Dizi appeared on the historical stage of solo performance, and its performance styles consisted of the Southern school and the Northern school. Today, the Northern school and the Southern school are two main styles of Dizi music. The Northern school is characterized as more lively, and ornamentally technical with extensive use of different types of tricky fingering techniques and tonguing. The Northern school is mainly played by Bangdi (Check out Hong Kong Chinese Orchestra's introductory video to the Bangdi in {\href{https://www.youtube.com/watch?v=zJjfFqat\_oA}{https://www.youtube.com/watch?v=zJjfFqat\_oA}}). By contrast, the Southern school is more melodic. It can display the soft features of the Jiangnan region. The representative playing techniques of the Souther school include trills, upper acciaccatura, and so on. The Southern school is mainly played by Qudi (Check out Hong Kong Chinese Orchestra's introductory video to the Qudi in \href{https://www.youtube.com/watch?v=HjU5ssXvYQA}{https://www.youtube.com/watch?v=HjU5ssXvYQA}). 

The rest of this paper is organized as follows. Related work is first shown in section 2. Data preparation is described in section 3. Section 4 and section 5 introduce deconstruction and reconstruction, respectively. Conclusions and some future work are given in section 6. 

\section{Related Work}

The first part of related work is data collection. The collected data of Chinese music consists of two forms: audio and symbolic scores. In terms of audio database establishment, Liang et al. \cite{liang2019constructing} built a database that includes Chinese musical instrument audio. Wang et al. \cite{wang2019cbf} collected a Dizi audio music dataset. In terms of symbolic music scores collection, Li et al. \cite{li2012automatic} collected a Gongchepu (which is the Chinese traditional musical notation) dataset. Li et al. \cite{guqin} collected a Guqin dataset.

The second part of related work is music deconstruction. Music deconstruction includes research about melody, music spectral characteristics, the correlation of different types of music genres, and so on. Wang et al. \cite{wang2020playing} did research about playing techniques recognization from the Dizi music spectrum. Yang et al. \cite{yang2014cross} did a quantitative study of vibrato to compare erhu music and violin music. 

The third part of related work is music reconstruction. Music reconstruction includes music generation, music synthesis, and so on. Luo et al. \cite{luo2020mg} used methods of deep learning to generate Chinese folk songs with specific styles. Dai et al. \cite{pipa} did music synthesis based on modeling of pipa playing techniques.

\section{Data Preparation}
\subsection{Data Collection}
Chinese musical instruments are usually recorded in the form of numbered musical notation. Therefore, we propose a general method to collect numbered musical scores. Although this method is applied in Dizi in this paper, it can be extended easily to other Chinese musical instruments which are also recorded in numbered musical notation. This method consists of three steps: making a new font, typing, and transformation.

\begin{figure}[htbp]
	\centering
	\subfigure[The overview of the font design]{
		\includegraphics[width=6.8cm]{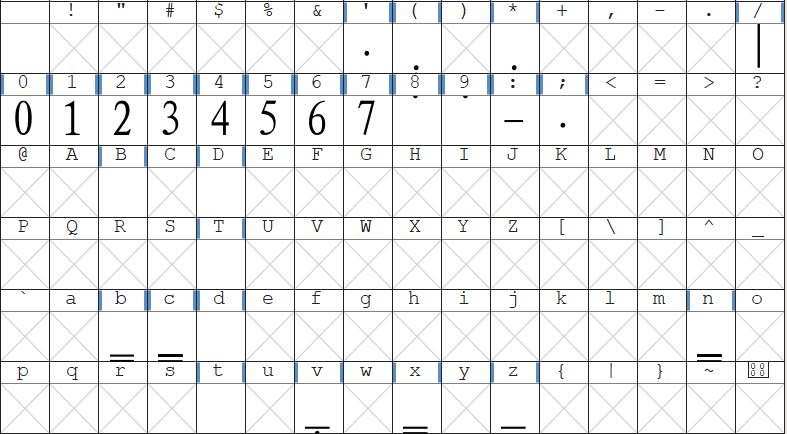}
	}
	\qquad
	\subfigure[Making lower acciaccatura]{
		\includegraphics[width=2.5cm]{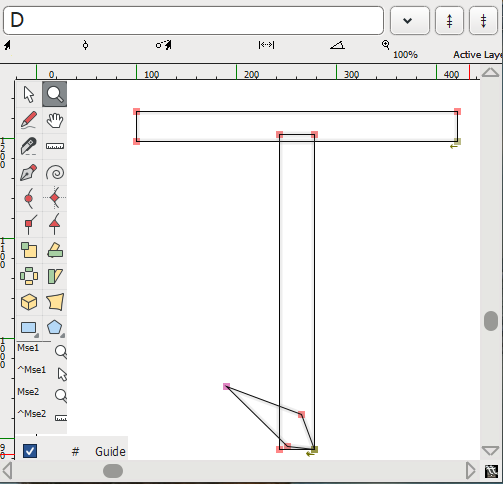}
	}
	\label{fig:sub2}
	\caption{The method of making DiziFont.ttf }
	\label{fig:font design}
\end{figure}

First, we used the open-source software FontForge to make a new font. We call the new font DiziFont.ttf. The method of making DiziFont.ttf can be seen in Fig. \ref{fig:font design}. The left subfigure shows the overview of the font design. Some keys in the keyboard correspond to some symbols in the numbered musical notation. For example, the method of making lower acciaccatura is shown in the right subfigure. The symbol of lower acciaccatura consists of three polygons of different shapes and sizes, and it corresponds to the capital letter D on the keyboard.

\begin{figure}[htbp]
	\centerline{
		\includegraphics[width=8cm]{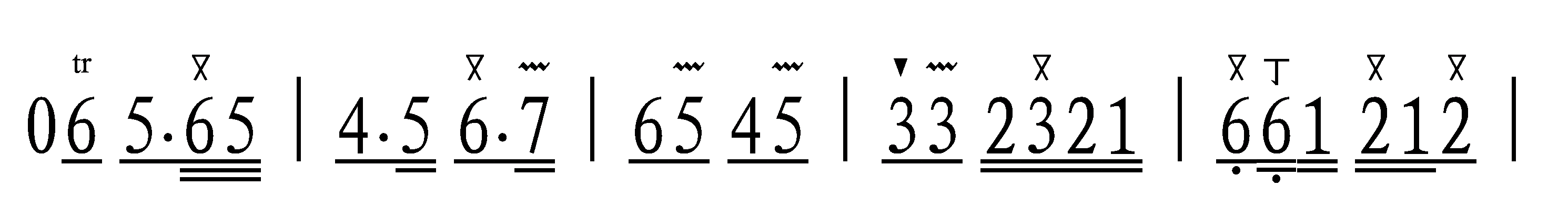}}
	\caption{A typing example}
	\label{fig:An example}
\end{figure}

Second, we applied the font of DiziFont.ttf into Microsoft word to type in numbered musical notation. A piece of example can be seen in Fig. \ref{fig:An example}. From this figure, we can see that the digitized numbered musical notation looks the same as on paper, which is very intuitive. The paper sheet music we used comes from \cite {lizhen2003} and \cite{Dizi1993}. The original typing files are stored in Docx files.  

\begin{figure}[htbp]
	\centerline{
		\includegraphics[width=12.0cm]{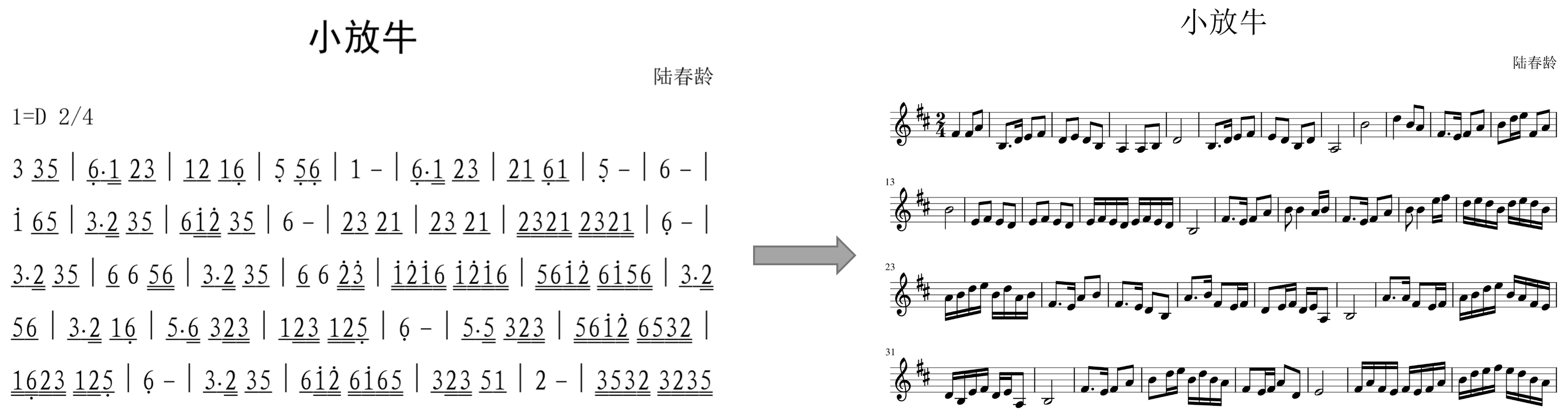}}
	\caption{A transform example}
	\label{fig:A transform example}
\end{figure}

Third, we wrote a program using the music21 toolkit \cite{cuthbert2010music21} to transform the Docx file into the MusicXML file. Although it is intuitive to record the numbered musical notation in the Docx file using our self-made font, it is not standardized. MusicXML file is not only a more standardized store form but also can be displayed in the form of staff using some software such as MuseScore. A transform example from the Docx file to the MusicXML file can be seen in Fig. \ref{fig:A transform example}. In practice, the playing technique symbols are too complicated which brings us great difficulties to process them. Therefore, we used an unintuitive but very simple way to record and process playing techniques. In this way, playing techniques are represented as lyrics to add to staff scores.

Using this above method, we collect a Dizi dataset both in Docx files and MusicXML files (which is also to say, numbered musical scores and staff scores).

\subsection{Data Statistics}

Up to now, we have recorded 28 Dizi songs, which include 19413 notes in total. The dataset is still being continuously expanded. In terms of style, these songs cover the Southern school, the Northern school, and so on. In this paper, we use the Northern school data and the Southern school data to analyze. The total number of notes used in this paper is 12925, where 7320 notes for the Northern school and 5605 notes for the Southern school.

\subsection{Data Representation}
\begin{figure}[htbp]
	\centerline{
		\includegraphics[width=4.68cm]{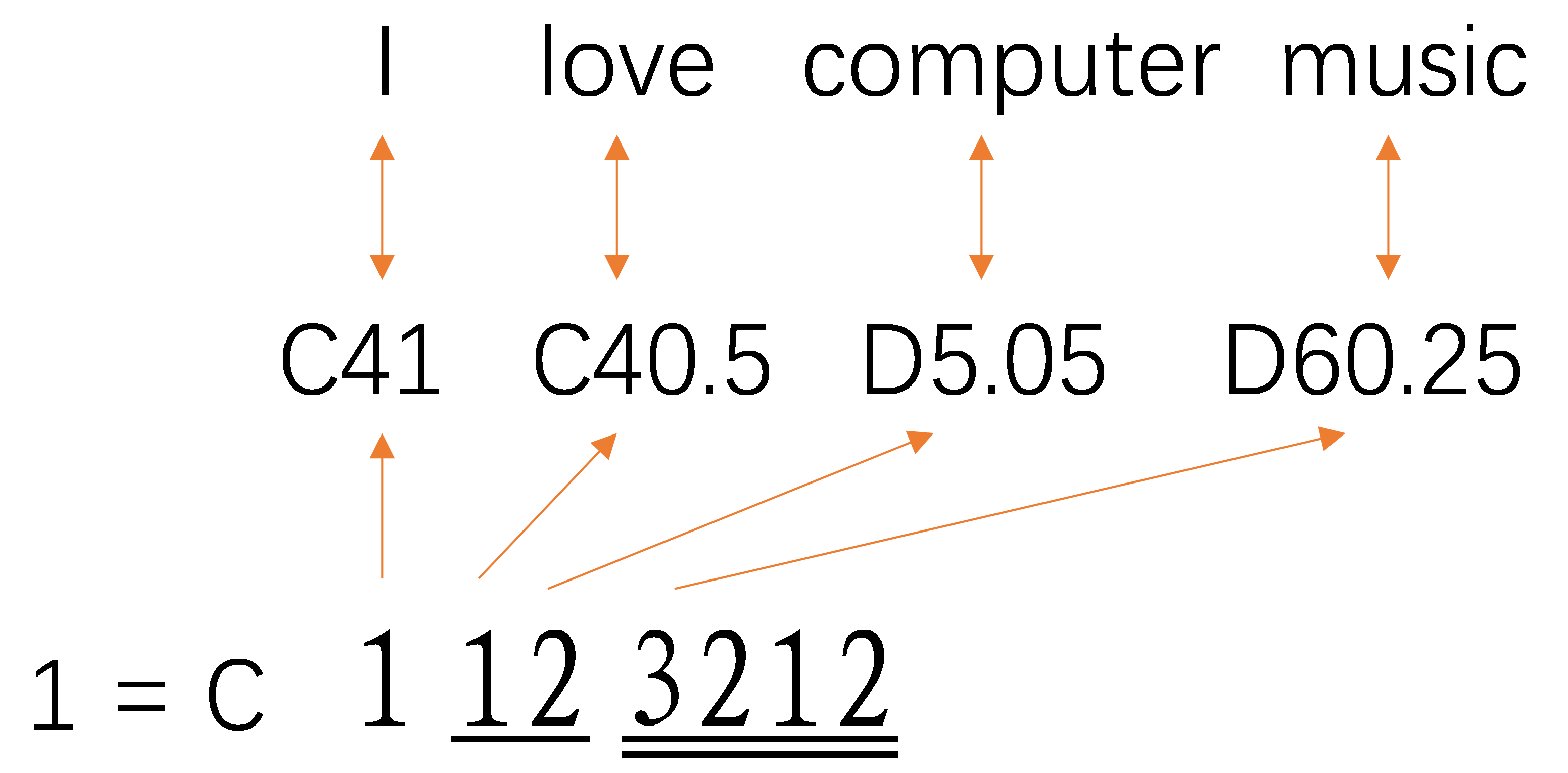}}
	\caption{A data representation example}
	\label{fig:A data representation example}
\end{figure}

In this paper, we focus on Dizi music of symbolic representation, so each note can be seen as a word just like in natural language processing. Each note consists of two features: the pitch and the duration. We use the chromatic scale to measure the pitch and quarter length to measure the duration. An example is shown in Fig. \ref{fig:A data representation example}. It can be seen how the processing of symbolic music is related to natural language processing. It shows a note sequence (represented in the form of numbered musical notation) in C major. The quarter note Do in C major has the pitch of C4 and the duration of 1 quarter length. The pitch and the duration can be spliced together and expressed as C41, and the other notes are the same.

\section{Deconstruction}

Deconstruction is used to find some information about features from music, which is done through data mining. For most kinds of music, the melody is quite an important feature, so we first did the melody deconstruction. For Dizi music, playing techniques are also important. Different styles have different representative playing techniques, so we then did the playing techniques deconstruction. In this paper, deconstruction is seen as a typical classification task. 

\subsection{Melody Deconstrction}

Melody deconstruction is used to classify melodies of the Northern school and the Southern school, which can also be seen as a general text classification task. We first cut music into many pieces, each of whose length is 4 measures, then used data preprocessing techniques include Bag-of-Words, Term Frequency Inverse Document Frequency
(TF-IDF) \cite{tfidf}, Continuous Bag-of-Words (CBOW) \cite{mikolov2013efficient} and Skip-Gram \cite{mikolov2013efficient} to process these data, finally sent these preprocessed-data to some usual machine learning models include Support Vector Machines (SVM) \cite{hearst1998support}, long-short-term memory (LSTM) \cite{graves2013speech} and Text Convolutional Neural Network (TextCNN) \cite{kim2014convolutional}. Experiments were done under 10-fold cross validation. Recall and F1-socre were used to evaluate experiment results. 

The results are shown in Tab. \ref{tab:obev1}. From this table, we can see that LSTM+TF-IDF and LSTM+Bag-of-Words get the relative best results. Besides, we find that the results of using LSTM are better than using TextCNN in total, which is an interesting result. In the short-text classification task, TextCNN has been proved to perform better than LSTM in many tasks, but the Dizi melody classification is not so. We think it is because that TextCNN can only do convolution operation during a small range, but LSTM can memory longer data. Compared with text data, music more depends on long memorized data. Grasping partial features not global features are difficult to recognize melody style.

\begin{table}[!htbp]
	\centering
	\caption{Melody deconstruction results}
	\begin{tabular}{c||c|c|c|c|c|c|c|c}
		\hline
		& \multicolumn{2}{c|}{Bag-of-Words} & \multicolumn{2}{c|}{TF-IDF} & \multicolumn{2}{c|}{CBOW} & \multicolumn{2}{c}{Skip-Gram}\\
		\hline
		Model & Recall & F1-score & Recall & F1-score & Recall & F1-score & Recall & F1-score\\
		\hline
		SVM & 97.89 & 89.96& 98.77 & 95.82 & 97.96 & 92.21 & 97.81 & 92.23 \\
		LSTM & 98.41 & 97.45 & 98.10 & 97.74 & 95.35 & 94.16 & 93.87 & 94.72\\
		CNN & 89.66 & 89.28 & 96.11 & 95.40 & 93.76 & 88.71 & 87.83 & 82.39\\
		\hline
	\end{tabular}
	\label{tab:obev1}
\end{table}

\subsection{Playing Techniques Deconstrction}
Playing techniques deconstruction is seen as a special classification task, tagging task. The tagging task is discussed between the observation sequence and the state sequence. In this playing technique deconstruction, each kind of playing technique is seen as a state and each note is seen as an observation. We first cut music into pieces which one of whose length is 4 measures, then used random word embedding to preprocess data, finally sent these preprocessed-data to some tagging models. Experiments were done under 10-fold cross-validation in the dataset of the Northern school and the dataset of the Southern school, respectively. Accuracy and oov (out-of-vocabulary) accuracy were used to evaluate experimental results.

Besides, Special instructions are needed regarding the tagging model used. Besides these usual tagging models include Conditional Random Fields (CRF) \cite{lafferty2001conditional}, bidirectional LSTM (BILSTM) \cite{graves2013hybrid} and BILSTM with a CRF layer (BILSTM-CRF) \cite{huang2015bidirectional}, we also used the model proposed in \cite{xie2020symbolic}, which combines a general tagging model and logic rules. As the general tagging model we used is BILSTM, we call this model BILSTM-RULES. 

The total experiment results are shown in Tab. \ref{tab:obev2}. We can see that BILSTM-CRF achieves the highest accuracy among two datasets of different styles, while BILSTM-RULES achieves the highest oov accuracy.

\begin{table}[!htbp]
	\centering
	\caption{Playing techniques deconstruction results. In this table, N represents the Northern school and S represents the Southern school.}
	\begin{tabular}{c|c|c|c|c}
		\hline
		& \multicolumn{2}{c|}{Accuracy} & \multicolumn{2}{c}{Oov accuracy} \\
		\hline
		Model & N & S & N & S\\
		\hline
		CRF & 68.98 & 84.42  & 39.44 & 63.03\\
		BILSTM & 69.76 & 84.03& 61.29 & 88.26 \\
		BILSTM-CRF & 74.71 & 87.59 & 43.54 & 85.53\\
		BISLTM-RULES & 69.23 & 84.12 & 61.95 & 88.79\\
		\hline
	\end{tabular}
	\label{tab:obev2}
\end{table}

\begin{figure}[htbp]
	\centering
	\subfigure[Composed by Xianwei Jiang]{
		\includegraphics[width=5.66cm]{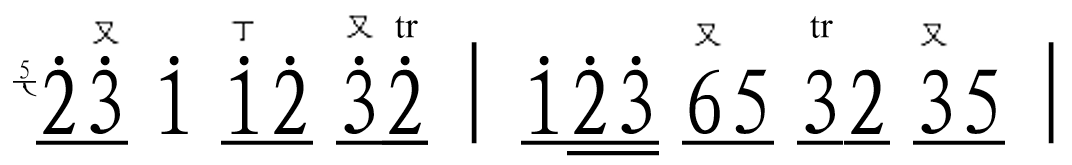}
	}
	\quad
	\subfigure[Composed by Xunfa Yu]{
		\includegraphics[width=5.66cm]{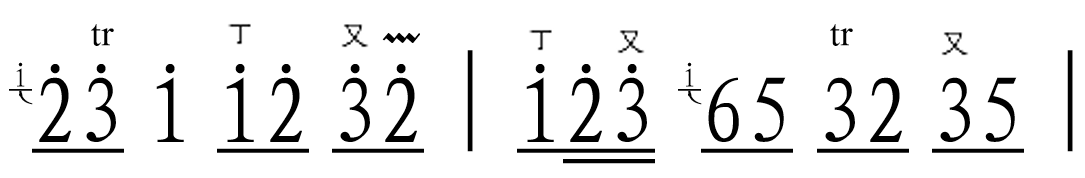}
	}
	\subfigure[Original techniques]{
		\includegraphics[width=3.87cm]{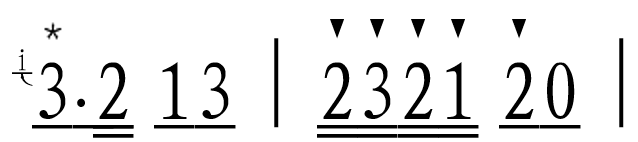}
	}
	\quad
	\subfigure[Gotten by BILSTM-CRF]{
		\includegraphics[width=3.87cm]{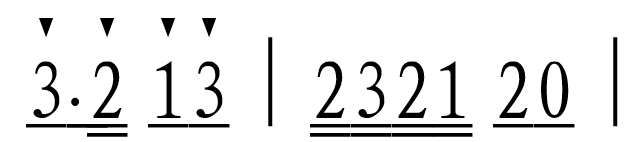}
	}
	\caption{Different techniques tags for the same notes}
	\label{fig:gusuxing}
\end{figure}

Besides, we find an interesting phenome, that is, not the same as the original data label does not mean it does not meet a certain style. Although composers from the same school can have different playing techniques tagging ways for the same music. For example, the first two subfigures in \ref{fig:gusuxing} show an excerpt of \emph{Song of Soochow}. Xunfa Yu and Xianwei Jiang are both from the Southern school, but they tagging the music differently. In our playing techniques deconstruction, there are also some similar examples of happening. An example is shown in the last two subfigures in \ref{fig:gusuxing}, it shows an excerpt of \emph{Busy Delivering Harvest}. We can see that playing techniques are different between the original and the generated. But from the human perspective, the playing techniques generated by BILSTM-CRF still meet the original style, the Northern school. As tonguing (which is represented as a triangle in numbered musical notation) is a typical playing technique from the Northern school.

\section{Reconstruction}

After deconstruction, we can reconstruct new music using deconstruction results. In this paper, we use music style transfer as a reconstruction example, to show how to get new music from deconstruction results. There are two steps of reconstruction (music style transfer) in this paper: melody reconstruction (melody transfer) and playing techniques reconstruction (playing techniques transfer). 

\subsection{Melody Reconstrction}

The first step is melody transfer. Given a piece of music with a specific style, then add some changes to this piece of music. The changes include the following four kinds:

\begin{itemize}
	\item Changing a note one or two chromatic semitones higher than the old one.
	\item Changing a note one or two chromatic semitones lower than the old one.
	\item Splitting one note into several notes, these notes' duration include $\left( \frac{1}{2}, \frac{1}{2} \right)$, $\left( \frac{1}{3}, \frac{2}{3} \right)$ , $\left( \frac{2}{3}, \frac{1}{3} \right)$ and $\left( \frac{1}{3}, \frac{1}{3}, \frac{1}{3} \right)$  
	\item Joining two successive notes into one note whose duration is the sum of these two notes’ duration.
	
\end{itemize}

Then, this piece of music with changes was sent to LSTM with the pre-processing technique of TF-IDF (which has been proved to perform well in Section 4.1). If the prediction label of classification is the same as its initial label (keeping the content of original music), and the probability of prediction is smaller than the old (It is closer to the style of target music), then these changes could be retained, else be dropped. The above process was repeated until the specified number of iterations is reached.

\subsection{Playing Techniques Reconstrction}

The second step is playing techniques transfer. We used models in Section 4.2 Which achieve the highest accuracy (BILSTM-CRF) and oov accuracy (BILSTM-RULES), respectively, to do playing techniques transfer. For example, if you want to transfer a piece of music that belongs to the Northern school style, into the style of the Southern school, you only need to apply the trained Southern school's playing techniques tagging models to the piece of music.

\subsection{Reconstruction Result}

\begin{figure}[!htbp]
	\centering
	\subfigure[Original melodies adding generated playing techniques]{
		\includegraphics[width=10.8cm]{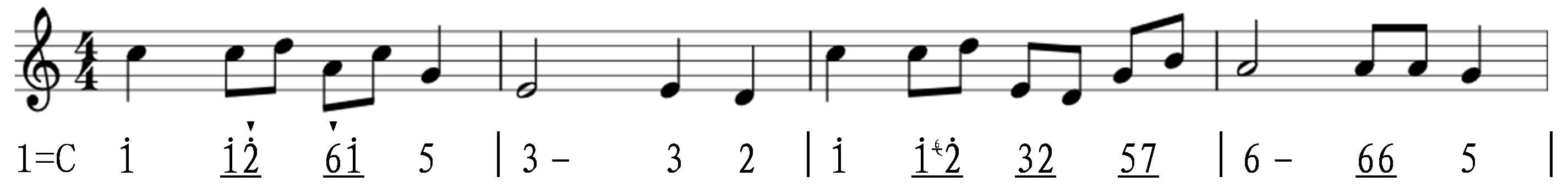}
	}
	\quad
	\subfigure[Generated melodies after 20 iterations adding generated playing techniques]{
		\includegraphics[width=10.8cm]{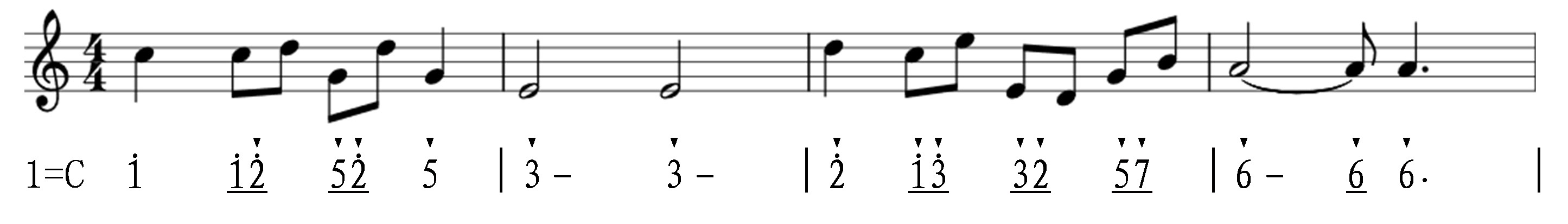}
	}
	\quad
	\subfigure[Generated melodies after 60 iterations adding generated playing techniques]{
		\includegraphics[width=10.8cm]{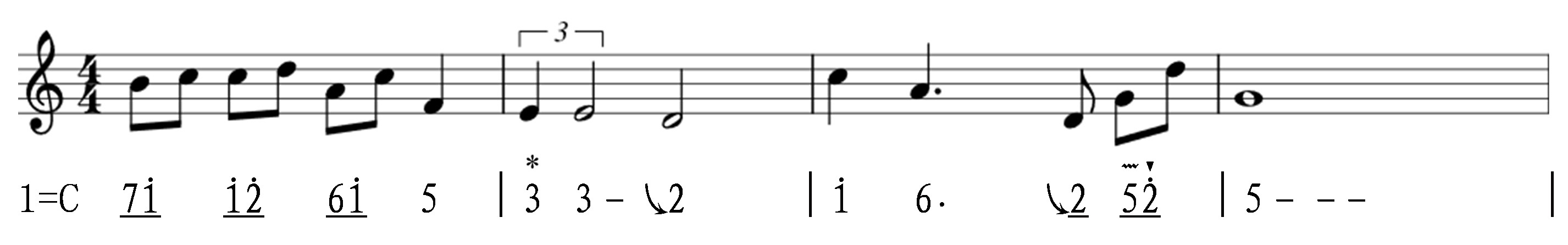}
	}
	\caption{A transfer example from the the Southern school to the Northern school. The original melodies come from \emph{Song of Soochow}.}
	\label{fig:transfer examples}
\end{figure}

After doing melody reconstruction and playing techniques reconstruction, the final reconstruction results can be gotten. A reconstruction example that shows the music style transfer result from the Southern school to the Northern school can be seen in Fig. \ref{fig:transfer examples}. In each subfigure, the first row represents notes in the staff, the second row represents notes and playing techniques in the form of numbered musical notation. It can be seen that with the increase of iteration number, more kinds of playing techniques related to the Northern school (NT) is been generated. Adding the original melodies with playing techniques using our models, only tonguing appears two times. After 20 iterations of melodies, tonguing appears more times than the original melodies. After 60 iterations, typical playing techniques with the Northern school features like flutter tonguing, portamento appear, too.

\subsection{Reconstruction Evaluation}
After getting reconstruction results, we did the evaluation. In the reconstruction evaluation, we set four subtasks in total: the Northern school to the Southern school (N2S), the Southern school to the Northern school (S2N), the other school to the Northern school (O2N), the other school to the Southern school (O2S). After getting reconstruction results in the form of symbolic music, we played them in Dizi to get audios. We made a questionnaire to do an evaluation. There are 35 participants in total, and all of them have related music background. The score of evaluation is from 1 to 10. A higher score means a more thorough reconstruction. For the music style transfer task, it is not only needed to transfer to the target style, but also needed to maintain the original content, so it is a good result to get an upper-middle score.

The evaluation results are shown in Tab. \ref{tab:subj}. In general, we can see that for all four subtasks, the score of using both melody reconstruction and playing techniques reconstruction, is higher than the score of using only melody reconstruction. Besides, we can see that BILSTM-CRF performs better when the Northern school is seen as the music transfer target, while BILSTM-RULES performs better when the Southern school is seen as the transfer target.

\begin{table}[!htbp]
	\centering
	\caption{Results of audience evaluation. In this table, BILSTM-CRF and BILSTM-RULES are both methods of playing techniques reconstruction}
	\begin{tabular}{c|c|c|c|c}
		\hline
		Method & S2N & N2S & O2N & O2S \\
		\hline
		Only melody reconstrction & 4.79 & 6.00 & 5.82 & 5.50 \\
		Melody reconstrction + BILSTM-CRF & 5.41 & 6.36 & 6.79 & 6.29\\ 
		Melody reconstrction + BILSTM-RULES & 5.18 & 6.50 & 6.15 & 6.65 \\
		\hline
	\end{tabular}
	\label{tab:subj}
\end{table}

\section{Conclusions and Future Work}

In this paper, we proposed a general collection method of Chinese music and collected a Dizi symbolic dataset. Using some machine learning methods, we trained some classification models which laid the foundation for music reconstruction and found some interesting phenomena. We also gave a reconstruction example about music style transfer, where audience tests were done to do the evaluation. 

Future work includes maintenance and expansion of the Dizi dataset, as well as broader and in-depth applications and research based on this dataset.

\section{Acknowledgement}
Supported by MOE (Ministry of Education in China) Youth Project of Humanities and Social Sciences, No.19YJCZH084
%
%
%
\bibliographystyle{splncs03}
\bibliography{mybib}

\end{document}